\documentclass{article}
\usepackage[utf8]{inputenc}
\usepackage[english]{babel}
\usepackage[T1]{fontenc}
\usepackage{authblk}

\usepackage[a4paper,top=3cm,bottom=2cm,left=3cm,right=3cm,marginparwidth=1.75cm]{geometry}

\usepackage{amsmath}
\usepackage{graphicx}
\usepackage{float}
\usepackage[colorinlistoftodos]{todonotes}
\usepackage[colorlinks=true, allcolors=blue]{hyperref}
\usepackage{soul} 

\usepackage{natbib}
\title{A note on metapopulation models}
\author[1]{Diepreye Ayabina$^\dag$}
\author[2]{Hasan Sevil$^\dag$}
\author[2]{Adam Kleczkowski}
\author[2,3]{M. Gabriela M. Gomes*}
\affil[1]{Liverpool School of Tropical Medicine, Liverpool, United Kingdom}
\affil[2]{Department of Mathematics and Statistics, University of Strathclyde, Glasgow, United Kingdom}
\affil[3]{NOVA School of Science and Technology, Centre for Mathematics and Applications (NOVA MATH), Caparica, Portugal}
\date{}
\begin{document}

\maketitle
\def\thefootnote{\dag}\footnotetext{These authors contributed equally to this work}
\def\thefootnote{*}\footnotetext{Corresponding author: gabriela.gomes@strath.ac.uk}

\begin{abstract}
Metapopulation models are commonly used in ecology, evolution, and epidemiology. These models usually entail homogeneity assumptions within patches and study networks of migration between patches to generate insights into conservation of species, differentiation of populations, and persistence of infectious diseases. Here, focusing on infectious disease epidemiology, we take a complementary approach and study the effects of individual variation within patches while neglecting any form of disease transmission between patches. Consistently with previous work on single populations, we show how metapopulation models that neglect in-patch heterogeneity also underestimate basic reproduction numbers ($\mathcal{R}_{0}$) and the effort required to control or eliminate infectious diseases by uniform interventions. We then go beyond this confirmatory result and introduce a scheme to infer distributions of individual susceptibility or exposure to infection based on suitable stratifications of a population into patches. We apply the resulting metapopulation models to a simple case study of the COVID-19 pandemic.
\end{abstract}

\section{Introduction}

Based on how they handle population structure, epidemiological models fall into a spectrum ranging from simple models that assume that the population is made up of a homogeneous group of randomly mixing individuals, to models that allow individual-level heterogeneities to be taken into account in disease dynamics. There are several sources of individual-level variability which affect contact rate, susceptibility, infectiousness, recovery rate, among others, that shape disease dynamics. Analysis of simple models that do not take this variability into account is tractable and insightful but may deviate from reality in significant ways \citep{Bansal2007, Corder2020, Gomes2022, Gomes2024}. On the other hand, although individual-based models incorporate the most detailed descriptions of individuals and societies \citep{Eubank2004, Ajelli2010, Beyrer2012, Megiddo2014, Moghadas2020, Kerr2021}, they are often intractable and require a significant amount of resources to develop \citep{Willem2017}. At an intermediate level of complexity are metapopulation models, which partition the entire population into a number of sub-populations or patches according to some form of geosocial structure \citep{Levins1969, Hess1996, Lloyd1996, Cross2005}. It is usually assumed that individuals are homogeneous and mix at random within patches and attention is concentrated on aspects of connectivity between patches that enable species or disease persistence. 

There is a growing body of literature on metapopulation models of infectious diseases that focus on taking explicitly into account the movement of individuals \citep{Keeling2002, Grais2003, Ruan2006} and a coupled force of infection that depends on interactions between local populations \citep{Park2002, Vazquez2007}. In order to specify these models, several distributional assumptions and parameters are needed and it is often desirable to estimate all or some of these from observations \citep{Ball2015}. Consequently, for human diseases, much effort has been invested into the collection and analysis of human mobility data \citep{Colizza2008, Meloni2011, Wesolowski2012, Tizzoni2014, Panigutti2017, Rubrichi2018, Wu2020, Chinazzi2020}.

Here we focus on deviations from assumptions of homogeneity within patches. These could have a sociologic (in terms of contact/interaction) or biological  (susceptibility or infectiousness) basis \citep{Becker1983}. Individual-level heterogeneity is increasingly documented to influence the transmission of infectious diseases \citep{Willem2017}. For intuition, consider a population with individual-level heterogeneity in risk (susceptibility or exposure) to an infectious disease. The first individuals to be infected are those with higher risk, leaving behind a pool of susceptible individuals with an average risk that decreases over time, by a process of selection. This decelerates transmission in relation to a homogeneous population with the same initial mean risk. Conversely, in order to sustain a given level of infection, a heterogeneous population requires a larger basic reproduction number and offers more attrition to uniform interventions \citep{Gomes2024}. We begin by showing how these insights, which were previously developed for single populations, extend systematically to metapopulation models. We then proceed to introduce an analytical scheme to infer distributions of individual susceptibility or exposure to infection based on suitable stratifications of a population into patches. Finally, we apply the resulting metapopulation models to a simple case study of the COVID-19 pandemic. We also discuss the incorporation of these insights and approaches in statistical schemes to enable the identification of unobserved variation, as illustrated in \citep{Gomes2019, Corder2020} for endemic, and \citep{Gomes2022} for epidemic, infectious diseases.

 \section{A metapopulation model with in-patch heterogeneity}
 
We consider a metapopulation model with $n$ patches (or $n$ populations). Disease transmission is modelled with Susceptible-Infectious (SI) or Susceptible-Infectious-Recovered (SIR) models within each patch. We populate the patches with individuals drawn from a common pool with a distribution of susceptibility or connectivity which we discretise into $m$ risk classes. Hence, for each risk value $x_j$, for $1\leq j\leq m$, we have
\begin{eqnarray}
    Q(x_j)&=&\sum_{i=1}^n \frac{N_i}{N} q_i(x_j),
\end{eqnarray}
where $N$ is size of the metapopulation, $N_i$, $1\leq i\leq n$, is the size of population $i$, $Q(x_j)$ is the proportion of the metapopulation that consists of individuals with risk value $x_j$, and $q_i(x_j)$, $1\leq i\leq n$ is proportion of population $i$ that consists of individuals with risk value $x_j$. By this construction, $Q$ and $q_i$, $1\leq i\leq n$, are probability mass functions. Figure 1 illustrates two different metapopulations created from the same common pool in a simple case with two patches and two risk classes. Given the tradition of metapopulation models with in-patch homogeneity, in this illustration the common pool was discretised into risk-class sizes that match the population sizes. This is theoretically convenient as it allows a systematic investigation of the effects of incorporating mixing between risk classes (starting from no mixing in the limit of homogeneous patches) and carries no loss of generality as any conceivably continuous risk distribution can be discretised into groups of any desirable sizes.

Throughout this paper we consider the number of risk classes equal to the number of patches (i.e., $m=n$). For theoretical convenience, in this section we also define risk-class sizes to match population sizes (i.e., $Q(x_i)=N_i/N$, for $i=1,\ldots ,n$). This condition is relaxed in Section \ref{COVID}, where various quantities are informed by real data which had been previously stratified in ways that constrain the analysis. Another feature of our models is that rather than introducing patch-specific effective contact rates as usually done in homogeneous-patch metapopulation models, we let transmissibility within the various patches be differentiated by the respective risk distributions: $q_i(x)$, for $i=1,\ldots ,n$. We thus assume a single transmission coefficient, $\beta$, which applies to all patches, and recover a homogeneous-patch model when individuals are distributed in a specific manner among patches. This homogeneous-patch model serves as a special case to which other compositions are compared to study the effects of in-patch heterogeneity.

The model equations are
 \begin{eqnarray}
\label{odes}
\begin{array}{l}
 \displaystyle \frac{S_{ij}}{dt} = \mu q_{i}(x_j) -(\lambda_i x_j + \mu)S_{ij} \\ \\
 \displaystyle \frac{I_{ij}}{dt} = \lambda_i x_j S_{ij} - (\gamma+\mu) I_{ij}
 \end{array}
\end{eqnarray}

 \noindent where $S_{ij}$ ($I_{ij}$) is the proportion of population $i$ which has risk level $j$ and is susceptible (infectious), and hence $\sum_j S_{ij}+I_{ij}=1$, for $i=1,\ldots ,n$. Parameter $\mu$ is the rate of host population turnover (through births and deaths) and $\gamma$ is the rate of recovery from infection. Subscripts $i=1,\ldots ,n$ identify the patches and $j = 1,\ldots ,n$ the risk levels, with $x_1\leq \ldots\leq x_j\leq\ldots\leq x_n$ being the susceptibility or connectivity values of individuals in their respective risk levels in relation to the metapopulation mean, which we assume equal to $1$, i.e.,
 \begin{eqnarray}
\label{mean1}
\langle x\rangle =\sum_{j=1}^{n} Q(x_j)x_j=1.
 \end{eqnarray}
 The coefficient of variation for the metapopulation is given by
 \begin{eqnarray}
\label{cv1}
\frac{\sqrt{\langle (x-\langle x\rangle)^2\rangle}}{\langle x\rangle} =\sqrt{\sum_{j=1}^{n} Q(x_j)(x_j-1)^2}.
 \end{eqnarray}
There is a force of infection $\lambda_i$ for each patch which is defined as
 \begin{eqnarray}
\label{foi_sus}
\lambda_i = \beta \sum_{j=1}^{n} I_{ij}
 \end{eqnarray}
 in the case of heterogeneous susceptibility, and
 \begin{eqnarray}
\label{foi_con}
\lambda_i = \frac{\beta}{\langle x\rangle_i} \sum_{j=1}^{n} x_j I_{ij}=\frac{\beta}{\sum_{j=1}^{n} q_i(x_j) x_j} \sum_{j=1}^{n} x_j I_{ij}
 \end{eqnarray}
when heterogeneity is in connectivity.

In the absence of transmission between the patches (which we assume throughout this paper as this is not relevant for the results presented here), the metapopulation basic reproduction number is
 \begin{eqnarray}
\label{R0}
\mathcal{R}_0 = \sum_{i=1}^{n} \mathcal{R}_{0i},
 \end{eqnarray}
where $\mathcal{R}_{0i}$ are the patch-specific basic reproduction numbers. When heterogeneity is in susceptibility we have
 \begin{eqnarray}
\label{R0_sus}
\mathcal{R}_{0i} = \langle x \rangle_i \frac{\beta}{\gamma+\mu},
 \end{eqnarray}
 while in the case of heterogeneous connectivity the basic reproduction number is
 \begin{eqnarray}
\label{R0_con}
\mathcal{R}_{0i} = \frac{\langle x^2 \rangle_i}{\langle x \rangle_i} \frac{\beta}{\gamma+\mu},
 \end{eqnarray}
where $\langle x \rangle_i$ and $\langle x^2 \rangle_i$ denote the first and second moments of the risk distributions specific to each patch. It can be shown that in the case of heterogeneous susceptibility $\mathcal{R}_0 = \langle x \rangle \beta/(\gamma+\mu)$, whereas in the case of heterogeneous connectivity we typically have $\mathcal{R}_0 \neq (\langle x^2 \rangle/\langle x \rangle) (\beta/(\gamma+\mu))$. 

\begin{figure}[H]
\includegraphics[trim={0 1.5cm 0 3.5cm},clip,width=\textwidth]{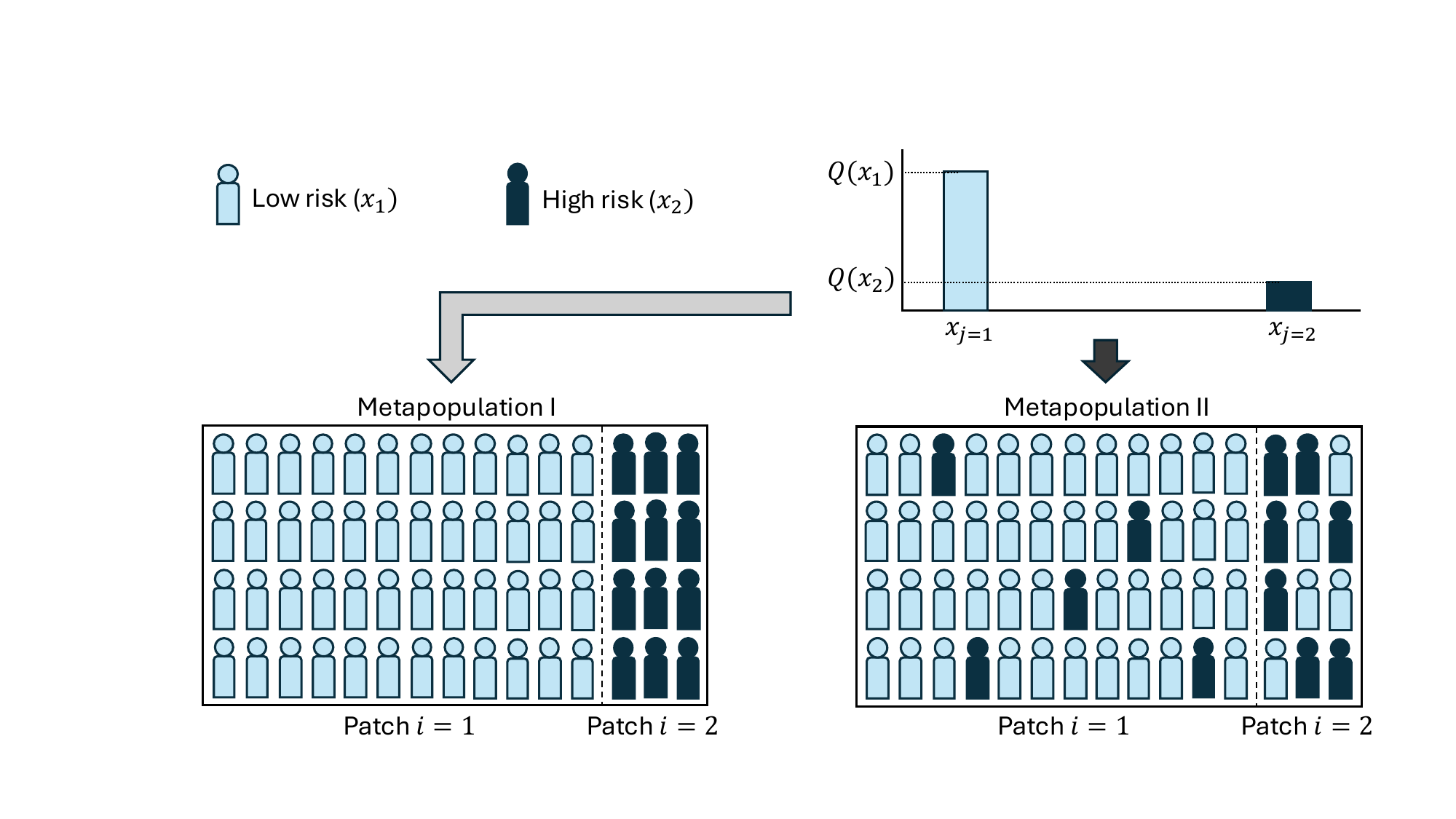}
\label{fig1}
\caption{The diagram represents two alternative metapopulations (I and II), each with two patches. In each case, patches are populated by individuals drawn from a given common pool containing low ($x_1$) and high ($x_2$) risk individuals in proportions $Q(x_1)$ and $Q(x_2)$, respectivelly. Each patch is thus characterized by its risk distribution. A single transmission coefficient is assumed for the entire metapopulation. }
\end{figure}

Given the construction above, all metapopulation models that are populated from the same pool have the same probability mass function $Q(x)$ and hence the same mean risk, but differ in how risk (and risk heterogeneity) is distributed among patches. To explore the effects of varying risk distribution and risk heterogeneity among patches, we conduct a series of interpolation analyses of metapopulation models with two patches and two risk values. For clarity of exposition, we refer to the patches as A and B (instead of $i=1$ and $i=2$) while the risk values remain indexed as 1 and 2 as represented above by $j$. For concreteness, we assume that $80\%$ of the metapopulation live in Patch A while the remaining $20\%$ reside in Patch B. Coincidentally, we also assume that $80\%$ of the individuals in the metapopulation are low risk while the remaining $20\%$ are high risk, and convention that Patch A always has lower risk that Patch B. 

With these assumptions, given the probability mass function for the metapopulation and the proportion of individuals in one of the risk classes for one of the patches, we can uniquely determine the probability mass functions for both patches. Hence, it is natural to adopt, for instance, the proportion of high-risk individuals in the lower-risk patch ($q_A(x_2)$) as a parameter to interpolate between two special scenarios: one where both patches are homogeneous ($q_A(x_2)=0$, and, consequently, $q_B(x_2)=1$); the other where patches are maximally heterogeneous and have identical risk distributions ($q_A(x_2)=q_B(x_2)=Q(x_2)$). All analyses are presented separately for SI and SIR model cases. The case $\gamma=0$ in system (\ref{odes}) gives the SI model, whereas $\mu=0$ corresponds to the SIR model without demography.

\subsection{Endemic (SI) case} \label{Sec:end}

Every realisation of the SI model ((\ref{odes}) with $\gamma=0$) supports an endemic equilibrium conditioned on $\beta$ being large enough. Throughout this section we assume a rate of population turnover (through births and deaths) compatible with a human population in a time scale of years ($\mu=1/80$ per year). 

\subsubsection{SI model with heterogeneous susceptibility} \label{Sec:endsus}

The main panel of Figure 2(a) shows the endemic prevalence of infection, according to model (\ref{odes}) with force of infection (\ref{foi_sus}), $\mathcal{R}_0$ (\ref{R0})-(\ref{R0_sus}), $Q(x_1)=N_A/N$, $Q(x_2)=N_B/N$, and parameterised by the proportion of the Patch A population (Population A) that consists of high-risk individuals ($q_A(x_2)$, represented as the horizontal axis). Recalling that $q_A$ and $q_B$ are probability mass functions we write the system as
 \begin{eqnarray}
\label{ode8}
\begin{array}{l}
\displaystyle \frac{dS_{A1}}{dt} = \mu q_A(x_1) -x_1\beta (I_{A1}+I_{A2}) S_{A1} - \mu S_{A1} \\ \\
\displaystyle \frac{dS_{A2}}{dt} = \mu q_A(x_2) -x_2\beta (I_{A1}+I_{A2}) S_{A2} - \mu S_{A2} \\ \\
\displaystyle \frac{dI_{A1}}{dt} = x_1\beta (I_{A1}+I_{A2}) S_{A1} - \mu I_{A1} \\ \\
\displaystyle \frac{dI_{A2}}{dt} = x_2\beta (I_{A1}+I_{A2}) S_{A2} - \mu I_{A2} \\ \\
\displaystyle \frac{dS_{B1}}{dt} = \mu q_B(x_1) -x_1\beta (I_{B1}+I_{B2}) S_{B1} - \mu S_{B1} \\ \\
\displaystyle \frac{dS_{B2}}{dt} = \mu q_B(x_2) -x_2\beta (I_{B1}+I_{B2}) S_{B2} - \mu S_{B2} \\ \\
\displaystyle \frac{dI_{B1}}{dt} = x_1\beta (I_{B1}+I_{B2}) S_{B1} - \mu I_{B1} \\ \\
\displaystyle \frac{dI_{B2}}{dt} = x_2\beta (I_{B1}+I_{B2}) S_{B2} - \mu I_{B2}.
 \end{array}
\end{eqnarray}
which for fixed parameter values has one solution with nonzero prevalence of infection when $\mathcal{R}_0>1$, as in the basic SI model \citep{AM1991}. The small panels on the left show the probability mass functions for both patches when $q_A(x_2)=0$, and those on the right when $q_A(x_2)=Q(x_2)$. The probability mass function for the metapopulation is assumed to have a coefficient of variation equal to 1, and hence $x_1=0.5$ and $x_2=3$ in the example shown. As expected, endemic prevalence increases in Patch A, and decreases in Patch B, as $q_A(x_2)$ increases. Less immediate, but not surprising, is that the metapopulation endemic equilibrium increases slightly as more high-risk individuals populate the larger patch. 

We assumed the metapopulation $\mathcal{R}_0=3$ for concreteness, and derived the corresponding $\beta$ from (\ref{R0})-(\ref{R0_sus}) which we applied throughout Figure 2. Recalling that $q_A$, $q_B$ and $Q$ are probability mass functions (hence $q_A(x_1)=1-q_A(x_2)$, $q_B(x_1)=1-q_B(x_2)$ and $Q(x_1)=1-Q(x_2)$), and that by construction $q_B(x_2)=1-Q(x_1)/Q(x_2)q_A(x_2)$, we derived values for distribution parameters $q_A(x_1)$, $Q(x_1)$, $q_B(x_1)$ and $q_B(x_2)$ given $Q(x_2)$ which we fixed at 0.2, and $q_A(x_2)$ which we varied within a range. Then, recalling that $Q$ has mean 1 we derived $x_1=(1-Q(x_2)x_2)/(1-Q(x_2)$. The prevalence curves were obtained by running system (\ref{ode8}) to equilibrium (using Matlab ode45) for values of $q_A(x_2)$ between $0$ and $0.2$ (horizontal axis), but they can also be derived by solving the system of 8 algebraic equations that results from setting the time-derivatives to zero.

Figure 2(b) simulates the transmission of infection, following identical introductions in both patches when $q_A(x_2)=0$ (left) and $q_A(x_2)=Q(x_2)$ (right). The patch-specific basic reproduction numbers are indicated by the corresponding growth curves. A small rate of transmission between patches might be added to these models if desired to assess sensitivity of the results to the extra parameter, but this has been subject of many studies and is not the focus of the present paper. Nevertheless, we have conducted such analyses (not shown) and the conclusions presented here were essentially unchanged.

\begin{figure}[H]
\includegraphics[trim={0 8cm 0 3.5cm},clip,width=\textwidth]{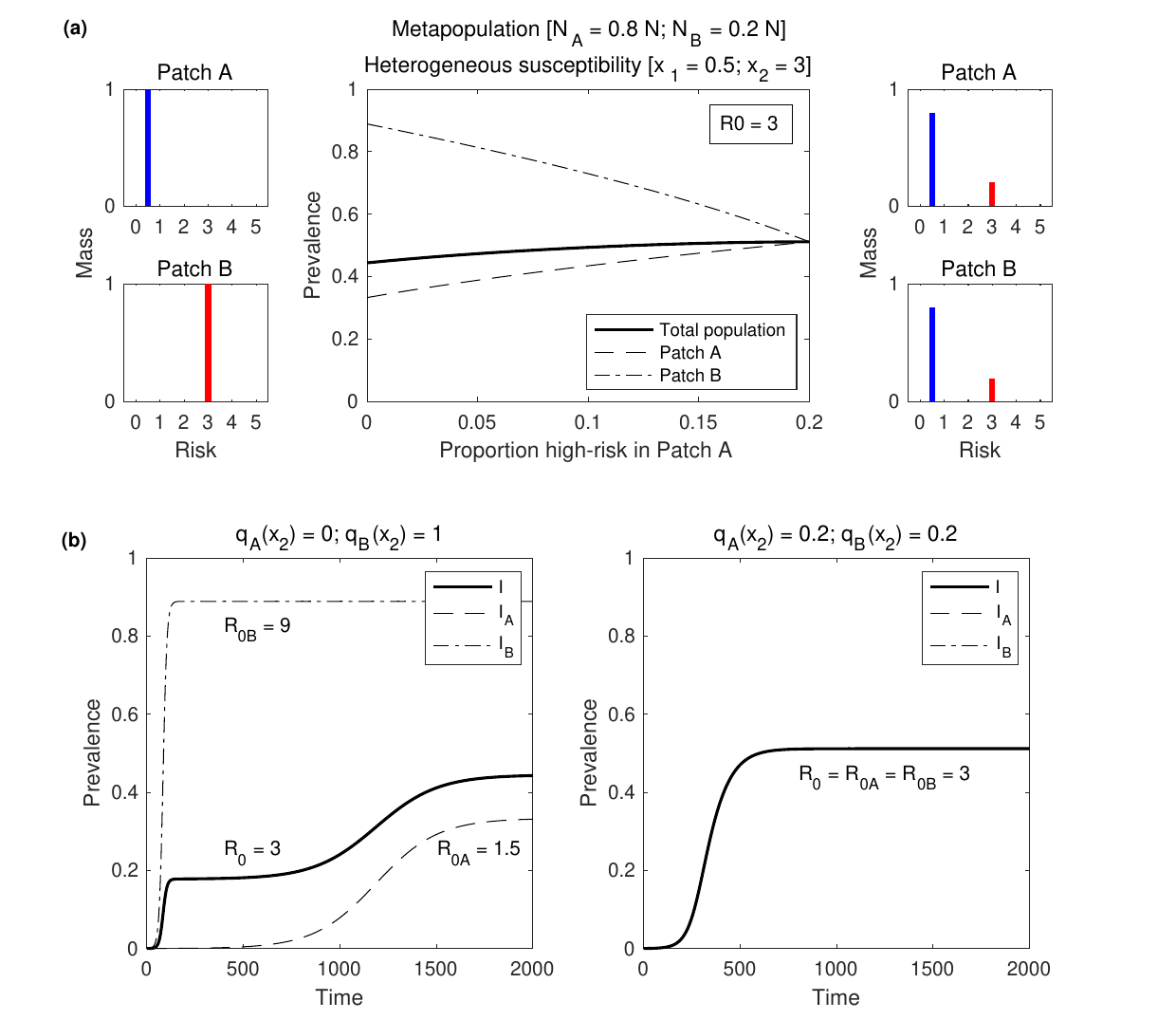}
\label{fig2}
\caption{Forward analysis of SI metapopulation model with heterogeneous susceptibility. (a) Endemic prevalence as a function of $q_A(x_2)$; (b) Simulated infection growth curves at the lower and higher admissible values for $q_A(x_2)$.}
\end{figure}

Following the straightforward interpolation above, we ask the more practical question of what might be the solution of an inverse problem where we are given the equilibrium prevalence in both patches and wish to infer $\mathcal{R}_0$. This inverse problem has a 1-dimensional array of solutions, each with a different $q_A(x_2)$ (and hence with a different coefficient of variation for the metapopulation probability mass function). Technically, we manipulate the system of 8 equilibrium equations associated with (\ref{ode8}) to obtain a system of 2 equations for $I_A$ and $I_B$
 \begin{eqnarray}
\label{equilibrium}
\begin{array}{l}
\displaystyle \frac{x_1q_A(x_1)}{\beta I_Ax_1+\mu} + \frac{x_2q_A(x_2)}{\beta I_Ax_2+\mu} = \frac{1}{\beta} \\ \\
\displaystyle \frac{x_1q_B(x_1)}{\beta I_Bx_1+\mu} + \frac{x_2q_B(x_2)}{\beta I_Bx_2+\mu} = \frac{1}{\beta}.
 \end{array}
\end{eqnarray}
We substituted expressions above for the distribution parameters $q_A(x_1)$, $Q(x_1)$, $q_B(x_1)$, $q_B(x_2)$ and $x_1$ (in terms of unknown $x_2$, and assumed $Q(x_2)$ fixed and $q_A(x_2)$ taking values within a range). We assumed $I_A=0.3$ and $I_B=0.5$ for concreteness, and solved system (\ref{equilibrium}) for $\beta$ and $x_2$ using Matlab function fsolve. From $\beta$ we then derived $\mathcal{R}_0$, and from $x_2$ we derived $x_1$ given the other distribution parameters.

Figure 3(a) shows the values obtained for the metapopulation $\mathcal{R}_0$ (solid line) as well as the basic reproduction numbers for the two patches ($\mathcal{R}_{0A}$ and $\mathcal{R}_{0B}$) as a function of the value assumed for the proportion of Population A that consists of high-risk individuals ($q_A(x_2)$). It is evident that all the basic reproduction numbers increase with $q_A(x_2)$. This is because higher $q_A(x_2)$ corresponds to more heterogeneous patches and, therefore, selective depletion of individuals at higher risk plays a larger role \citep{Gomes2019,Gomes2024}. In a randomly-mixing population (as in our patches), when individuals differ in their susceptibility to infection, the more susceptible tend to be infected first leaving uninfected a susceptible pool whose mean susceptibility decreases over time by selection. As a result, transmission decelerates and a higher $\mathcal{R}_0$ is required to attain the same endemic prevalence as in homogeneous populations with the same initial mean susceptibility. This is what causes the $\mathcal{R}_0$ curves to increase with $q_A(x_2)$.

\begin{figure}[H]
\includegraphics[trim={0 2cm 0 3cm},clip,width=\textwidth]{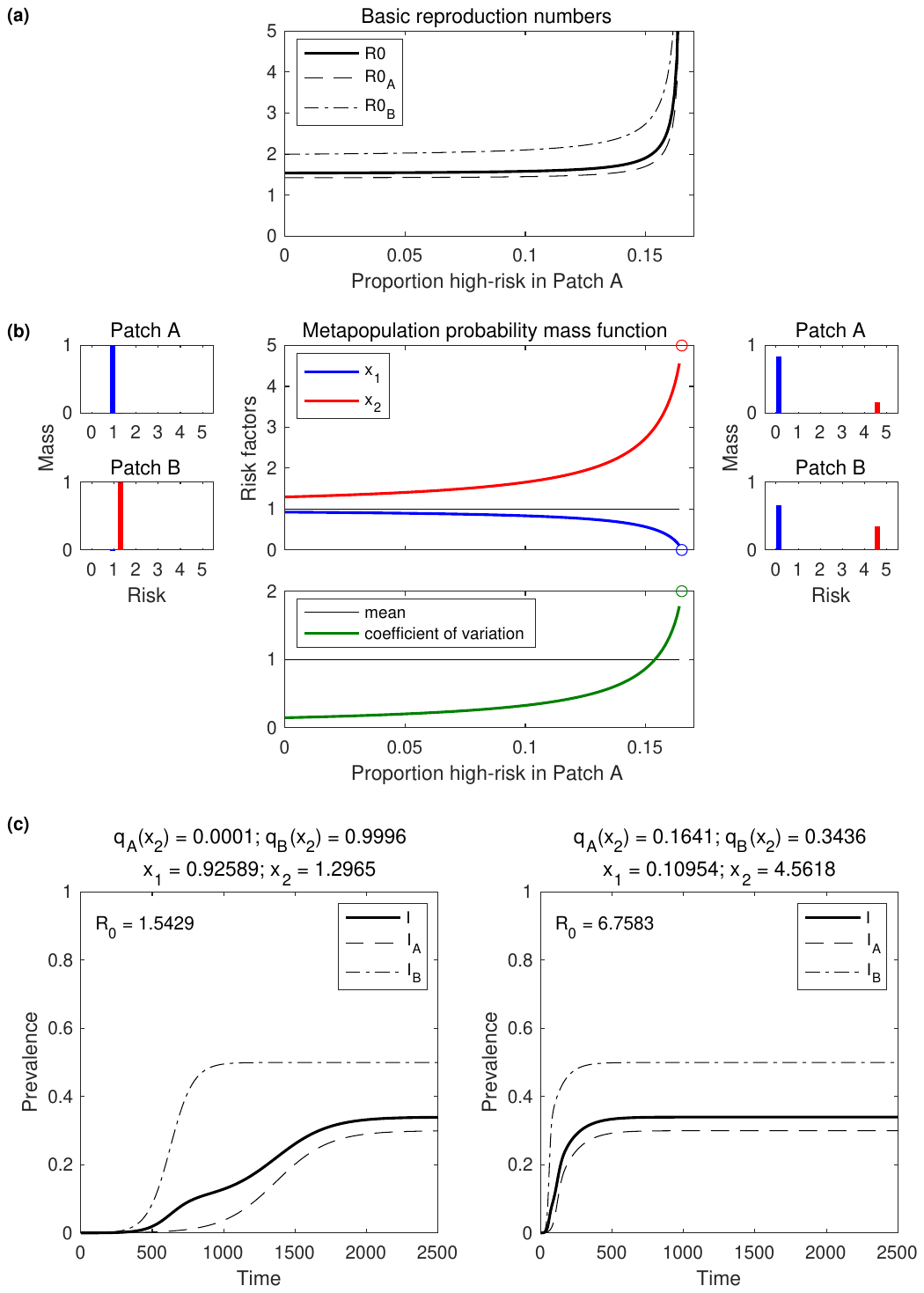}
\label{fig3}
\caption{Backward analysis of SI metapopulation model with heterogeneous susceptibility. $\mathcal{R}_0$ and $x_2$ were calculated given the prevalence of infection in both patches. (a) Basic reproduction numbers as a function of $q_A(x_2)$; (b) Properties of the risk distributions; (c) Simulated infection growth curves at low and high values of $q_A(x_2)$. Circles at the right end of the coloured lines indicate unattainable limits for this analysis.}
\end{figure}

Figure 3(b) provides more detail of how the probability mass functions vary as a function of $q_A(x_2)$. Not only do the proportions of low and high risk individuals become more balanced in both patches as $q_A(x_2)$ increases, but also the respective risk factors ($x_1$ and $x_2$) become more different. Both of these tendencies increase the opportunity for selective depletion and the consequent increase in basic reproduction numbers.

Finally, Figure 3(c) shows simulated growth curves of infection over time for a low and a high value of $q_A(x_2)$. The final endemic values are the same in the two cases because both are solutions of the same inverse problem, which differ in the value assumed for $q_A(x_2)$. To further demonstrate the importance of accounting for individual variation when solving inverse problems such as this, we apply an intervention that reduces the susceptibility of every individual by $50\%$ and show that the impact appears much greater when a model with more homogeneous in-patch transmission (and less selective depletion) is adopted (Figure 4).

\begin{figure}[H]
\includegraphics[trim={0 5cm 0 8.5cm},clip,width=\textwidth]{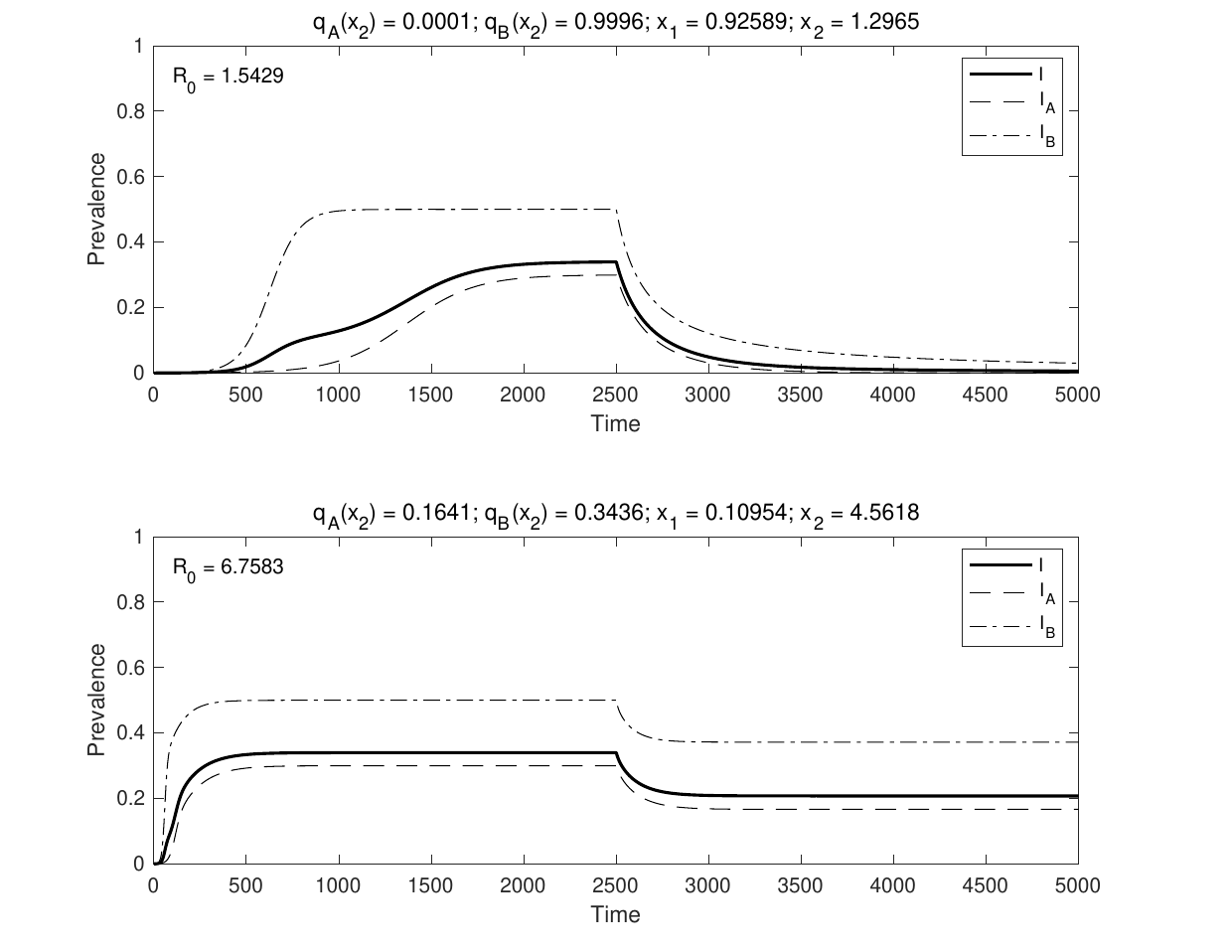}
\label{fig4}
\caption{Intervention impact according to SI metapopulation model with heterogeneous susceptibility. Simulated infection curves of Figure 3(c) prolonged with an intervention period in which susceptibility of all individuals was reduced by $50\%$.}
\end{figure}

\subsubsection{SI model with heterogeneous connectivity} \label{Sec:endcon}

In this section we replicate the analyses of Section \ref{Sec:endsus} for a model also given by (\ref{odes}) but with force of infection (\ref{foi_con}) and $\mathcal{R}_0$ (\ref{R0})-(\ref{R0_con}). The main panel of Figure 5(a) shows the endemic prevalence of infection as a function of $q_A(x_2)$, with the small panels on the left and right showing the probability mass functions when $q_A(x_2)=0$ and $q_A(x_2)=Q(x_2)$, respectively. In this example, endemic prevalence decreases with $q_A(x_2)$ in Patch B as in Section \ref{Sec:endsus}, while in Patch A there is no endemic equilibrium for low values of $q_A(x_2)$ as $\mathcal{R}_{0A}<1$ and prevalence increases with $q_A(x_2)$ from the point where $\mathcal{R}_{0A}$ crosses 1. The metapopulation endemic equilibrium then decreases slightly with $q_A(x_2)$ while $\mathcal{R}_{0A}<1$, and increases thereafter. 

In contrast with Section \ref{Sec:endsus}, the relationship between $\mathcal{R}_0$ and $\beta$ varies with the interpolation parameter $q_A(x_2)$ in this case (this can be verified by inspection on (\ref{R0})-(\ref{R0_con})). Hence, for illustrative completeness, while in Figure 5 we derive $\beta$ from $\mathcal{R}_0=3$ at $q_A(x_2)=Q(x_2)$ (when the patches are identical to the metapopulation), and fix the resulting value throughout the figure, in Supplementary Data Figure S1, we fix $\mathcal{R}_0=3$ and adopt $q_A(x_2)$-specific $\beta$ values. Interestingly, prevalence appears to decrease monotonically with $q_A(x_2)$ in the latter scenario.

\begin{figure}[H]
\includegraphics[trim={0 8cm 0 3.5cm},clip,width=\textwidth]{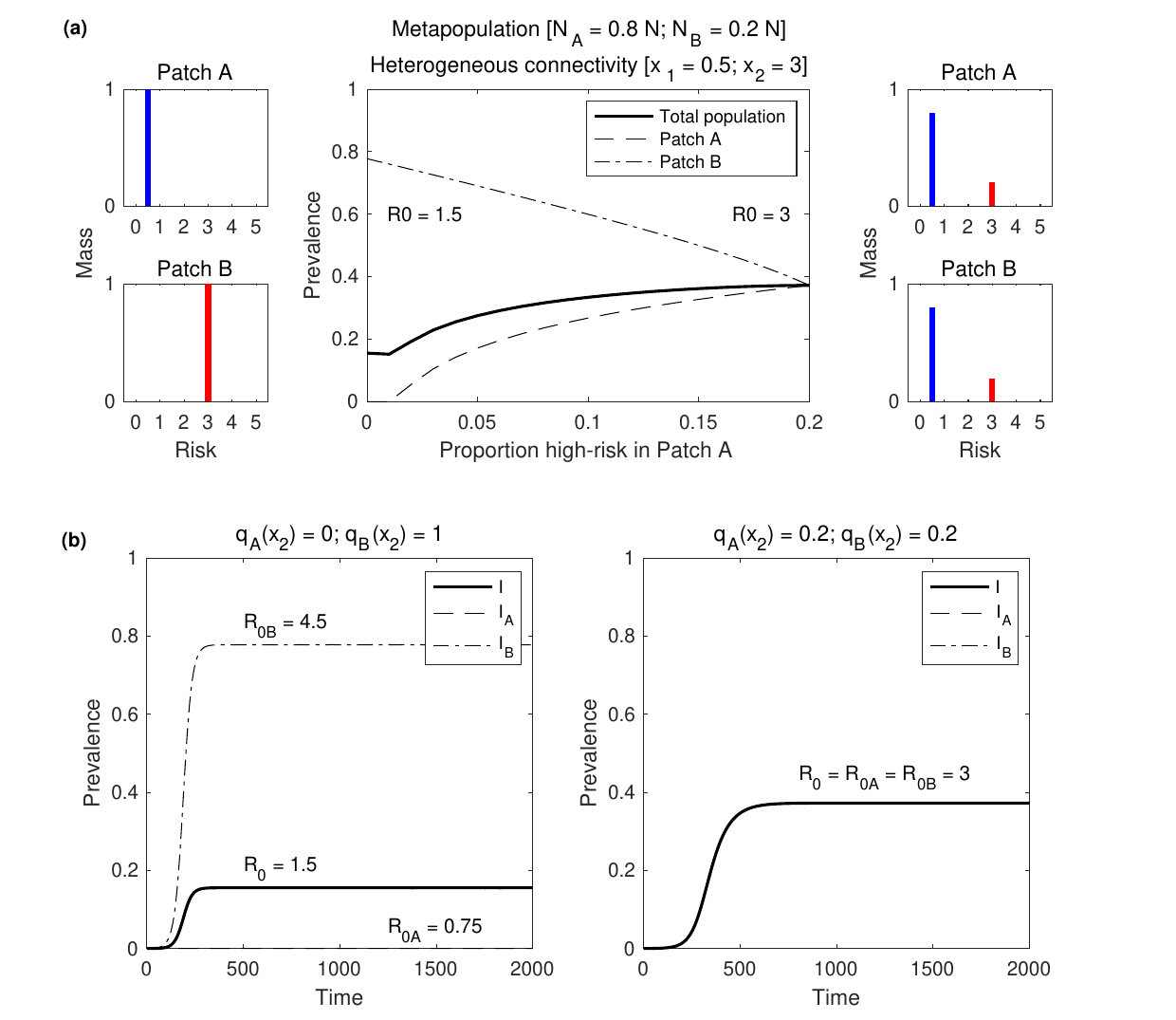}
\label{fig5}
\caption{Forward analysis of a SI metapopulation model with heterogeneous connectivity. (a) Endemic prevalence as a function of $q_A(x_2)$; (b) Simulated infection growth curves at the lower and higher admissible values for $q_A(x_2)$.}
\end{figure}

Figures 6 and 7 show the results of the inverse problem and intervention set up as in Section \ref{Sec:endsus}, with concordant results. In this case, however, we could not reduce the system of 8 algebraic equations describing the endemic equilibrium to only 2 equations, and solved the entire system 
 \begin{eqnarray}
\label{equilibrium_con}
\begin{array}{l}
\mu q_A(x_1) -x_1\beta (x_1I_{A1}+x_2I_{A2}) S_{A1} - \mu S_{A1} = 0 \\ \\
\mu q_A(x_2) -x_2\beta (x_1I_{A1}+x_2I_{A2}) S_{A2} - \mu S_{A2} = 0 \\ \\
x_1\beta (x_1I_{A1}+x_2I_{A2}) S_{A1} - \mu I_{A1} = 0 \\ \\
x_2\beta (x_1I_{A1}+x_2I_{A2}) S_{A2} - \mu I_{A2} = 0 \\ \\
\mu q_B(x_1) -x_1\beta (x_1I_{A1}+x_2I_{A2}) S_{B1} - \mu S_{B1} = 0 \\ \\
\mu q_B(x_2) -x_2\beta (x_1I_{A1}+x_2I_{A2}) S_{B2} - \mu S_{B2} = 0 \\ \\
x_1\beta (x_1I_{A1}+x_2I_{A2}) S_{B1} - \mu I_{B1} = 0 \\ \\
x_2\beta (x_1I_{A1}+x_2I_{A2}) S_{B2} - \mu I_{B2} = 0,
 \end{array}
\end{eqnarray}
using Matlab fsolve. The strategy of which parameters to fix, substitute and infer was the same as in Section \ref{Sec:endsus}.

\begin{figure}[H]
\includegraphics[trim={0 1cm 0 2cm},clip,width=\textwidth] {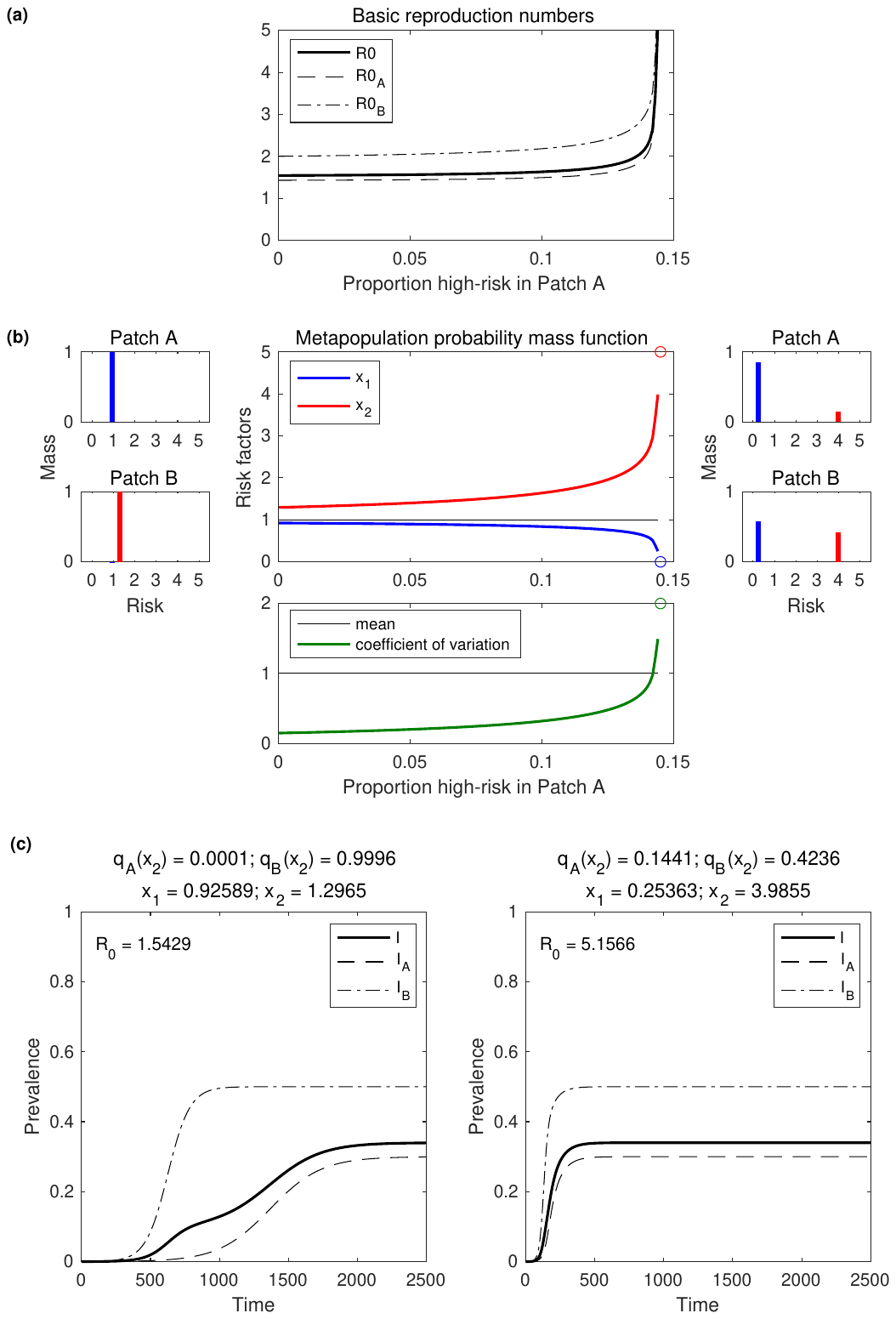}
\label{fig6}
\caption{Backward analysis of SI metapopulation model with heterogeneous connectivity. $\mathcal{R}_0$ and $x_2$ were calculated given the prevalence on infection in both patches. (a) Basic reproduction numbers as a function of $q_A(x_2)$; (b) Properties of the risk distributions; (c) Simulated infection growth curves at low and high values of $q_A(x_2)$. Circles at the right end of the coloured lines indicate unattainable limits for this analysis.}
\end{figure}

\begin{figure}[H]
\includegraphics[trim={0 5cm 0 8.5cm},clip,width=\textwidth] {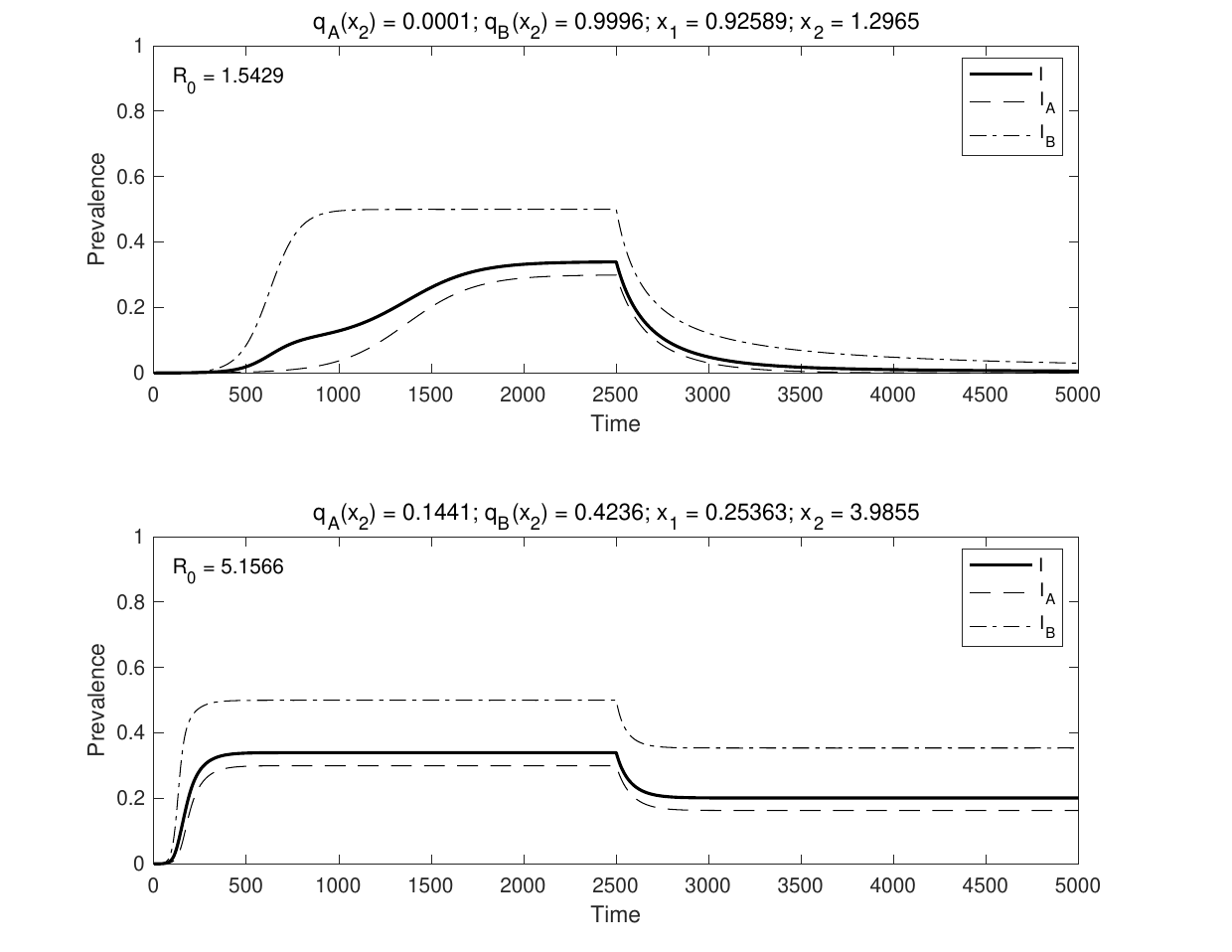}
\label{fig7}
\caption{Intervention impact according to SI metapopulation model with heterogeneous connectivity. Simulated infection curves of Figure 6(c) prolonged with an intervention period in which susceptibility of all individuals was reduced by $50\%$.}
\end{figure}

\subsection{Epidemic (SIR) case} \label{Sec:epi}

For each patch, a realisation of the SIR model ((\ref{odes}) with $\mu=0$) generates an epidemic curve conditioned on $\beta$ (or $\mathcal{R}_0$) being large enough. Throughout this section we assume a rate of recovery from infection compatible with many respiratory virus infection in a time scale of days ($\gamma=1/14$ per day). 

\subsubsection{SIR model with heterogeneous susceptibility} \label{Sec:episus}

The main panel of Figure 8(a) shows the epidemic final size, according to model (\ref{odes}) with force of infection (\ref{foi_sus}) and $\mathcal{R}_0$ (\ref{R0})-(\ref{R0_sus}), parameterised by $q_A(x_2)$. As in Section \ref{Sec:end}, the small panels on the left show the probability mass functions when $q_A(x_2)=0$, and those on the right when $q_A(x_2)=Q(x_2)$. The probability mass function for the metapopulation is assumed to have a coefficient of variation equal to 1, and hence $x_1=0.5$ and $x_2=3$. As expected, the epidemic final size increases in Patch A, and decreases in Patch B, as $q_A(x_2)$ increases. The metapopulation epidemic final size increases slightly as more high-risk individuals populate the larger patch, in agreement with results in Section \ref{Sec:endsus}. We assume the metapopulation $\mathcal{R}_0=3$, and derive the corresponding $\beta$ from (\ref{R0})-(\ref{R0_sus}) which we apply throughout the figure. Figure 8(b) simulates the transmission of infection, following identical introductions in both patches when $q_A(x_2)=0$ (left) and $q_A(x_2)=Q(x_2)$ (right). The patch-specific basic reproduction numbers are indicated by the corresponding growth curves.

Epidemic final sizes for this analysis were obtained implicitly from formulas derived in \citep{Katriel2012,MillerBMS2012}. In the case of our 2-patch SIR system (\ref{odes}) with heterogeneous susceptibility, the epidemic final sizes for the patches, $R_A(\infty)$ and $R_B(\infty)$, satisfy the relations
 \begin{eqnarray}
\label{efs_sus}
\begin{array}{l}
\displaystyle R_A(\infty) = 1-q_A(x_1)\exp{\left[-x_1\frac{\beta}{\gamma}R_A(\infty)\right]}-q_A(x_2)\exp{\left[-x_2\frac{\beta}{\gamma}R_A(\infty)\right]}\\ \\
\displaystyle R_B(\infty) = 1-q_B(x_1)\exp{\left[-x_1\frac{\beta}{\gamma}R_B(\infty)\right]}-q_B(x_2)\exp{\left[-x_2\frac{\beta}{\gamma}R_B(\infty)\right]}.
 \end{array}
 \end{eqnarray}

\begin{figure}[H]
\includegraphics[trim={0 8cm 0 3.5cm},clip,width=\textwidth]{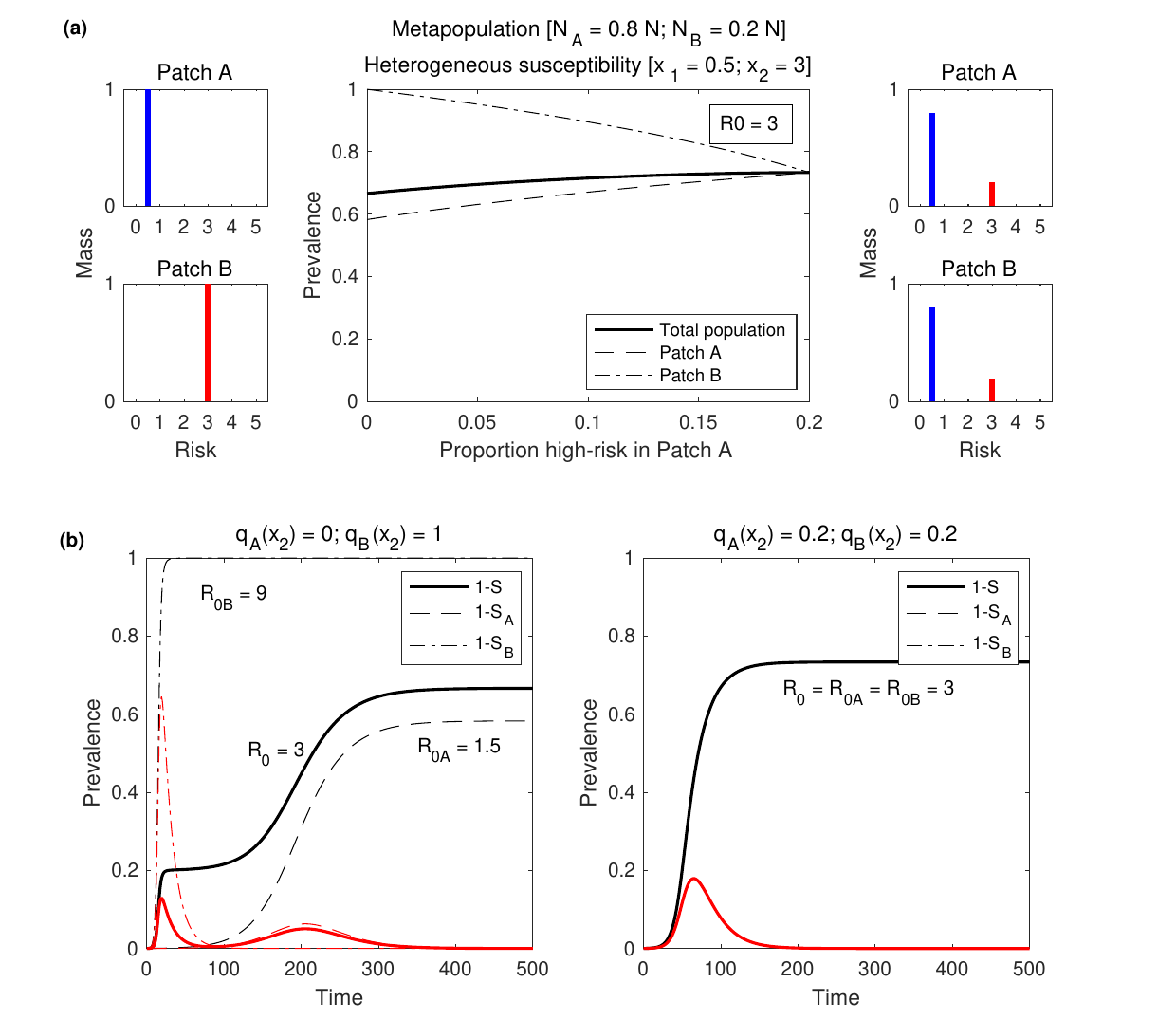}
\label{fig8}
\caption{Forward analysis of SIR metapopulation model with heterogeneous susceptibility. (a) Epidemic final size as a function of $q_A(x_2)$; (b) Simulated infection growth curves at the lower and higher admissible values for $q_A(x_2)$.}
\end{figure}

Formulas (\ref{efs_sus}) were also used to solve an inverse problem where an epidemic final size was assumed for each patch. As in previous sections, this inverse problem has a 1-dimensional array of solutions, each with a different $q_A(x_2)$ (and hence, a different coefficient of variation for the metapopulation probability mass function). As in Section \ref{Sec:end}, we substituted derived expressions the distribution parameters $q_A(x_1)$, $Q(x_1)$, $q_B(x_1)$, $q_B(x_2)$ and $x_1$ (in terms of unknown $x_2$, and assumed $Q(x_2)$ fixed and $q_A(x_2)$ taking values within a range). We assumed $R_A(\infty)=0.6$ and $R_B(\infty)=0.8$ for concreteness, and solved system (\ref{efs_sus}) for $\beta$ and $x_2$ using Matlab function fsolve. From $\beta$ we then derived $\mathcal{R}_0$, and from $x_2$ we derived $x_1$ given the other distribution parameters. Figure 9(a) shows the values obtained for the metapopulation $\mathcal{R}_0$ (solid line), as well as $\mathcal{R}_{0A}$ and $\mathcal{R}_{0B}$, parameterised by $q_A(x_2)$. All basic reproduction numbers increase with $q_A(x_2)$ due to selective depletion of susceptibles \citep{Gomes2022,Gomes2024}. 

Figure 9(c) shows simulated growth curves of infection over time for a low and a high value of $q_A(x_2)$. We then consider an intervention that would have reduced the susceptibility of every individual by $50\%$ preventively before the epidemic occurred, and show that the impact would appear greater in more homogeneous in-patch transmission scenarios (Figure S3).

\begin{figure}[H]
\includegraphics[trim={0 1cm 0 2cm},clip,width=\textwidth]{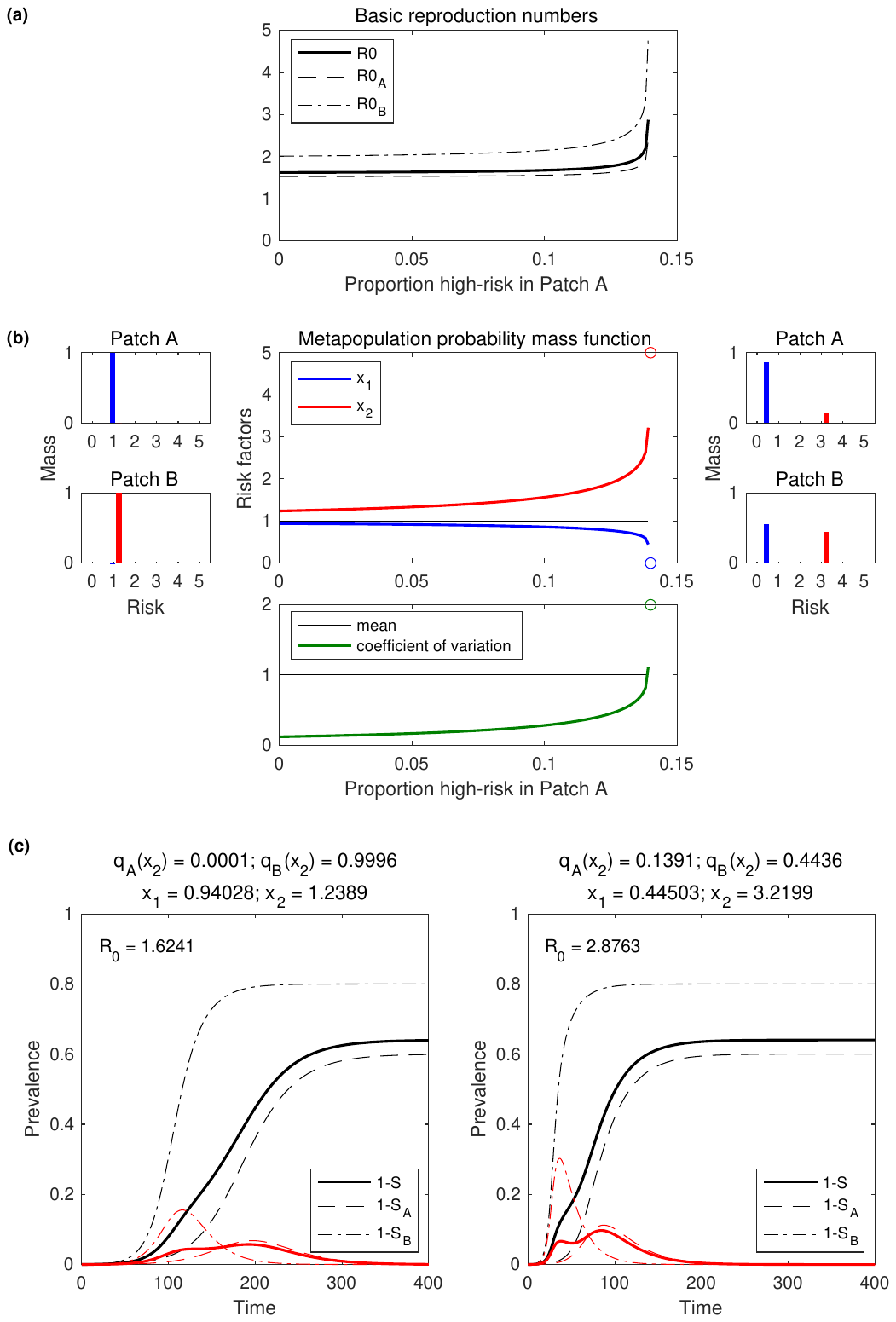}
\label{fig9}
\caption{Backward analysis of SIR metapopulation model with heterogeneous susceptibility. $\mathcal{R}_0$ and $x_2$ were calculated given the epidemic final sizes in both patches. (a) Basic reproduction numbers as a function of $q_A(x_2)$; (b) Properties of the risk distributions; (c) Simulated infection growth curves at low and high values of $q_A(x_2)$. Circles at the right end of the coloured lines indicate unattainable limits for this analysis.}
\end{figure}

\subsubsection{SIR model with heterogeneous connectivity} \label{Sec:epicon}

In this final case, we replicate the analyses of Section \ref{Sec:episus} for a model also given by (\ref{odes}) but with force of infection (\ref{foi_con}) and $\mathcal{R}_0$ (\ref{R0})-(\ref{R0_con}). The main panel of Figure 10(a) shows the epidemic final size as a function of $q_A(x_2)$. The epidemic final size decreases with $q_A(x_2)$ in Patch B, while in Patch A there is no epidemic for low values of $q_A(x_2)$ as $\mathcal{R}_{0A}<1$ and an epidemic whose size increases with $q_A(x_2)$ from the point where $\mathcal{R}_{0A}$ crosses 1. The metapopulation epidemic final size then decreases slightly with $q_A(x_2)$ while $\mathcal{R}_{0A}<1$, and increases thereafter. 

\begin{figure}[H]
\includegraphics[trim={0 8cm 0 3.5cm},clip,width=\textwidth]{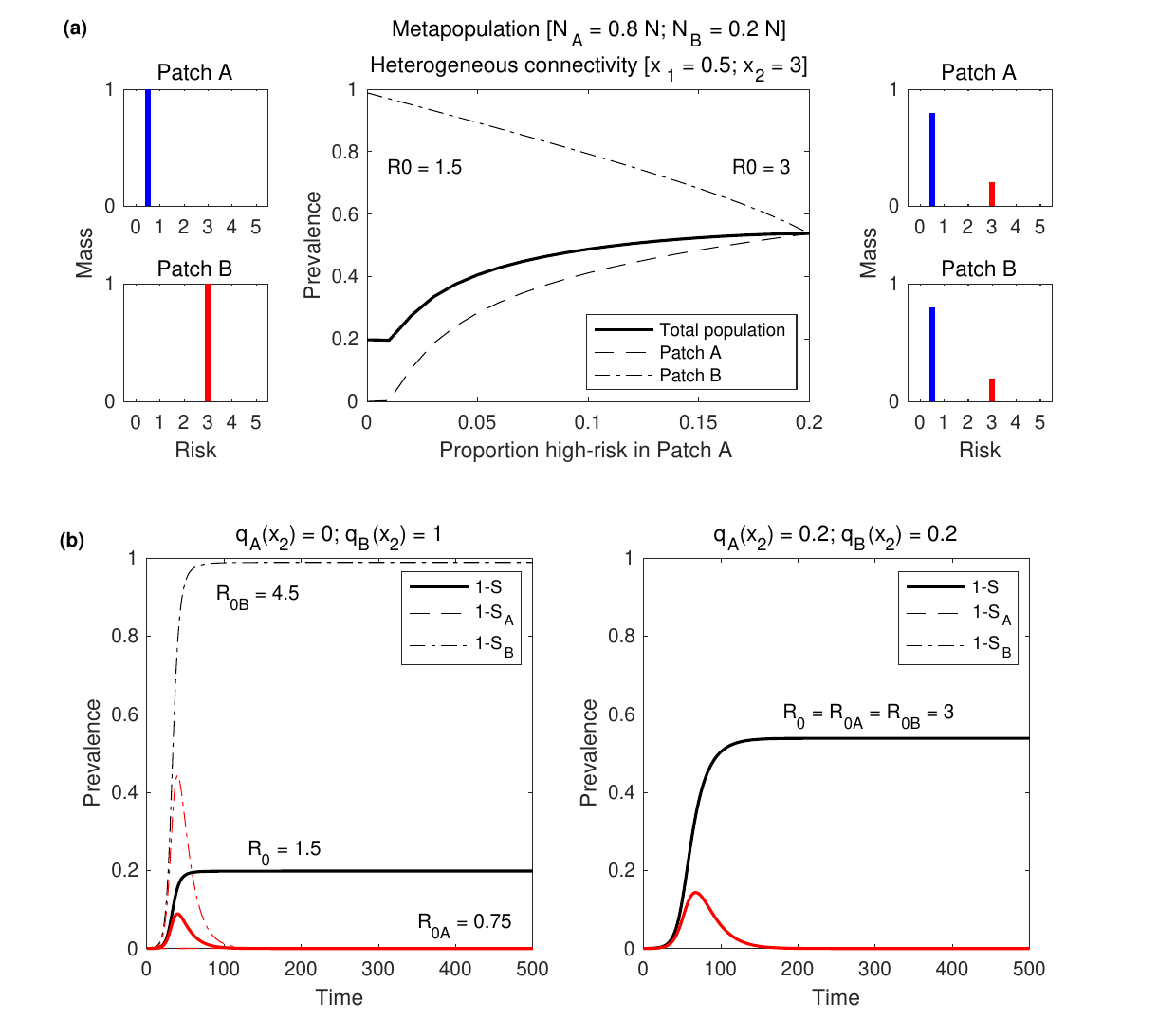}
\label{fig11}
\caption{Forward analysis of SIR metapopulation model with heterogeneous connectivity. (a) Epidemic final size as a function of $q_A(x_2)$; (b) Simulated infection growth curves at the lower and higher admissible values for $q_A(x_2)$.}
\end{figure}

As in Section \ref{Sec:endcon}, given the varying relation between $\mathcal{R}_0$ and $\beta$ in heterogeneous connectivity models, in Figure 10 we presents results for fixed $\beta$ (allowing $\mathcal{R}_0$ to vary), while in Supplementary Data Figure S2 we fix $\mathcal{R}_0=3$ and adopt $q_A(x_2)$-specific $\beta$ values. 

Epidemic final sizes were obtained implicitly from formulas derived in \citep{MillerJRSI2012}. In the case of our 2-patch SIR system (\ref{odes}) with heterogeneous connectivity, the epidemic final sizes for the patches, $R_A(\infty)$ and $R_B(\infty)$, satisfy the relations
 \begin{eqnarray}
\label{efs_con}
\begin{array}{l}
\displaystyle R_A(\infty) = 1-\frac{q_A(x_1)\theta_A^{x_1}+q_A(x_2)\theta_A^{x_2}}{q_A(x_1)x_1+q_A(x_2)x_2}\\ \\
\displaystyle R_B(\infty) = 1-\frac{q_B(x_1)\theta_B^{x_1}+q_B(x_2)\theta_B^{x_2}}{q_B(x_1)x_1+q_B(x_2)x_2},
 \end{array}
 \end{eqnarray}
 where $\theta_A$ and $\theta_B$ satisfy
\begin{eqnarray}
\label{ths_con}
\begin{array}{l}
\displaystyle \theta_A = \exp{\left[ -\frac{\beta}{\gamma}\left( 1- q_A(x_1)x_1\theta_A^{x_1}-q_A(x_2)x_2\theta_A^{x_2} \right) \right]}\\ \\
\displaystyle \theta_B = \exp{\left[ -\frac{\beta}{\gamma}\left( 1- q_B(x_1)x_1\theta_B^{x_1}-q_B(x_2)x_2\theta_B^{x_2} \right) \right]}.
 \end{array}
 \end{eqnarray}

Formulas (\ref{efs_con})-(\ref{ths_con}) were also used to solve an inverse problem where an epidemic final size was assumed for each patch. Strategy of which parameters to fix, substitute, infer (using Matlab fsolve) as in Section \ref{Sec:episus} with addition that in this case we also inferred $\theta_A$ and $\theta_B$. Figure 11(a) shows values obtained for $\mathcal{R}_0$, $\mathcal{R}_{0A}$, $\mathcal{R}_{0B}$, parameterised by $q_A(x_2)$. All basic reproduction numbers increase with $q_A(x_2)$ due to selective depletion of susceptibles. 

\begin{figure}[H]
\includegraphics[trim={0 1cm 0 2cm},clip,width=\textwidth]{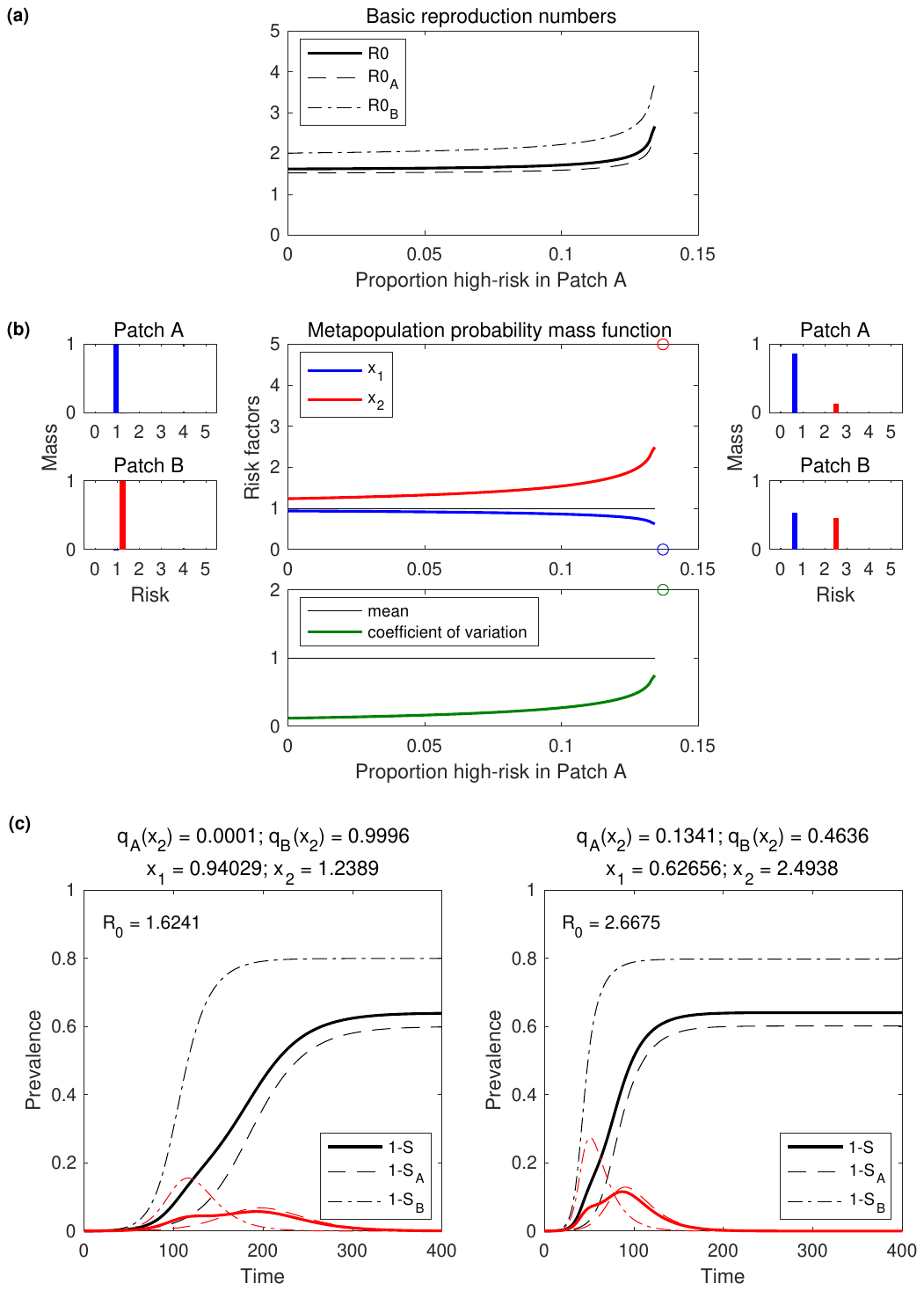}
\label{fig12}
\caption{Backward analysis of SIR metapopulation model with heterogeneous connectivity. $\mathcal{R}_0$ and $x_2$ were calculated given the epidemic final sizes in both patches. (a) Basic reproduction numbers as a function of $q_A(x_2)$; (b) Properties of the risk distributions; (c) Simulated infection growth curves at low and high values of $q_A(x_2)$. Circles at the right end of the coloured lines indicate unattainable limits for this analysis.}
\end{figure}

Figure 11(c) shows simulated growth curves of infection over time for a low and a high value of $q_A(x_2)$ and Figure S4 an intervention set up as in Section \ref{Sec:episus}, with concordant results.

\section{A case study of COVID-19 in two cities in Scotland} \label{COVID}

Here we present a real case study to illustrate the simplicity of how the models presented in previous sections might be applied to concrete situations to quantity the effects of specific risk factors in the acquisition of infectious diseases. As this is mainly for illustrative purposes, as we leave out modelling details that might be of interest to include in practice.

The coronavirus disease 2019 (COVID-19) pandemic affected different regions of the world differently, and these differences can be used to estimate how various factors affected the spread of the severe acute respiratory syndrome coronavirus 2 (SARS-CoV-2) using metapopulation models with in-patch heterogeneity. We used series of COVID-19 reported case data in Scotland extracted from \citep{GOVcases}, from the first cases in March 2020 until October 2023 (Supplementary Data Figure S5).

To describe risk distributions, we used socioeconomic deprivation data. The Scottish Index of Multiple Deprivation 2020 (SIMD) analyses data zones, which are small areas with similar population sizes ($784$ individuals on average), to identify areas where individuals are deprived in multiple ways in Scotland \citep{GOVdeprivation}. The index merges data from seven domains -- income, employment, health and disability, education, housing, crime, living environment -- to obtain
an overall relative measure of deprivation. The deciles of deprivation are derived by rating data zones nationwide according to these seven distinct variables. The first decile represents the $10\% $ most deprived zones, while the tenth decile represents the $10\% $ least deprived
areas. In order to quantify deprivation in each of our study regions, we use the percentage of data zones in that region whose SIMD is in the first 2 deciles.

We then apply the 2-patch SIR metapopulation models analysed in Section \ref{Sec:epi} to COVID-19 reported cases in the cities of Aberdeen and Dundee, using the first 2 SIMD deciles to quantify the proportions of high-risk individuals in each city (those who reside in data zones classified among the $20\% $ most deprived in Scotland). Using this scheme (see Supplementary Material for details) we find that Aberdeen (Patch A) has a high-risk proportion $q_A(x_2)=0.10$ while in Dundee (Patch B) the high-risk proportion is $q_B(x_2)=0.37$. From the cumulative case reports, we obtain final epidemic sizes as $R_A(\infty)=36\%$ in Aberdeen and $R_B(\infty)=40\%$ in Dundee. Using these data, we solve the inverse problems, which are set up by the final size equations (\ref{efs_sus}) when risk factors are assumed to reflect susceptibility, and (\ref{efs_con})-(\ref{ths_con}) when risk represents connectivity. In both scenarios, we obtain a metapopulation $\mathcal{R}_0=1.25$. For the metapopulation coefficient of variation, we obtain $0.047$ when heterogeneity is in susceptibility and $0.045$ when heterogeneity is in connectivity.

In \citep{Gomes2022}, a modelling study, which fitted Susceptible-Exposed-Infectious-Recovered (SEIR) models, with non-pharmaceutical interventions (NPIs) and continuous distributions of susceptibility or connectivity, to series of reported daily COVID-19 deaths in England and Scotland, estimated an $\mathcal{R}_0$ around $3$ and a coefficient variation of the order of 1. The substantial gaps between the estimates obtained in this section and the more thorough study conducted by \citep{Gomes2022} are not surprising, for several reasons.

First, the $\mathcal{R}_0$ estimated in this study is lower because case under-reporting and NPIs were not included in the models. Moreover, it is plausible that in more deprived regions, under-reporting would have been more severe (due to, for example, less frequent testing for COVID-19 \citep{Sevil2025}), reducing the difference in cases reported across regions and lowering estimates of risk variation. Second, \citep{Gomes2022} admitted that individual variation was unobserved and holistically estimated without specific factors attached, whereas here we use a deprivation index (SIMD) based on seven specific factors aggregated over data zones. It is plausible that the seven factors that SIMD is based on do not capture all the differences between individuals (e.g., age contributes to differences in susceptibility and connectivity and is not captured by deprivation factors, within the same deprivation decile some individuals are more socially active than others, etc). Both the omission of relevant factors and the aggregation over data zones are expected to cause the analysis presented here to underestimate risk variation. Future work should add realistic complexity to the metapopulation models studied here.

\section{Discussion}
The development of tractable models that capture the effects of individual variation in infectious disease dynamics is a topic of continued interested in mathematical epidemiology \citep{Ball1985, May1988, Morene2002, Miller2007, NOVOZHILOV2008, Gomes2019, Tkachenko2021, Zachreson2022, Gomes2022, Bootsma2024}, and several model simplifications have been compared to detailed individual-based models \citep{Ajelli2010, Keeling2010}. It has been repeatedly highlighted that models that under-represent the individuality of susceptibility and connectivity tend to exaggerate both the potency of disease transmission \citep{Ball1985, May1988, Miller2007, NOVOZHILOV2008, Ajelli2010, Keeling2010, Gomes2022, Gomes2024} and the expected impact of uniform control measures \citep{Gomes2024}. To close (or even reverse) this gap in the effectiveness of potential control measures between models that under-represent individual variation and reality it is of interest that individuals of high susceptibility or high social mixing are identifiable so that targeted interventions can be implemented.

Some modelling efforts admit that individual variation is not directly observable but it is inferable by fitting models with in-built distributions to suitable population data \citep{Halloran1996, Gomes2022, Gomes2024}. This approach is inspired in mathematical demography, econometrics, and survival analysis \citep{Vaupel1979, Singer1984, Hougaard1986, Aalen2015}. In cases where identifiability issues arise, these can usually be resolved in combination with metapopulation modelling (as in \citep{Gomes2022}, where COVID-19 models with distributed susceptibility or connectivity were simultaneously fitted to separate data series for England and Scotland, instead of a single series for the whole region). The main strength of this holistic approach is to capture individual variation entirely as indicated by the goodness of fits to population data. A weakness hinted above is that individual variation is treated as unobservable and hence available interventions cannot be targeted for maximal impact. The design of targeted interventions requires reductionist approaches where measurable risk factors are represented explicitly.

Here we introduce the scheme for such a reductionist approach. We consider SI and SIR models in metapopulations with $m$ risk classes and $n$ patches, where each patch might by inhabited by a mixture of individuals from various risk classes. We conduct systematic analyses for the case $m=n=2$ and apply the SIR version of the models to data from the cities of Aberdeen and Dundee, in Scotland, consisting of a socioeconomic deprivation index and COVID-19 cases between March 2020 and October 2023. 

For the systematic analysis, we use a parameter that interpolates between a scenario where mixing between risk classes is minimal (as different risk classes inhabit different patches) and a scenario where mixing is maximal. We then analyse SI and SIR models, each with heterogeneous susceptibility or connectivity (4 cases in total). SI models support an endemic equilibrium conditioned on $\mathcal{R}_0>1$. We assume fixed values for the endemic prevalence in each patch, as these are often the type of data available in real systems, and solve the inverse problem to infer the corresponding metapopulation $\mathcal{R}_0$ and the coefficient of variation (CV) of the metapopulation risk distribution, as a function of the interpolation parameter. In the case of SIR models, these support an epidemic conditioned on $\mathcal{R}_0>1$. We conduct a similar analysis but the inverse problem is set up to match epidemic final sizes, rather than endemic equilibria. We consistently find that both $\mathcal{R}_0$ and CV increase as the mixing parameter increases, a trend that we attribute to selective depletion of the most susceptible or most connected individuals \citep{Gomes2024}.

In the analysis of COVID-19 in Aberdeen and Dundee, we use a socioeconomic deprivation index -- the Scottish Index of Multiple Deprivation 2020 (SIMD) -- to determine whether individuals live in one of the $20\% $ most deprived data zones or not. For epidemic final sizes, we considered the cumulative number of cases reported by 01 October 2023 in each city. These are crude simplifications that we make here to illustrate how we envisage the models to be applied and further developed, rather than to thoroughly study COVID-19. With these assumptions and simplifications we estimated $\mathcal{R}_0\approx 1.25$ and CV$\approx 0.05$. In \citep{Gomes2022}, the authors had estimated $\mathcal{R}_0\approx 3$ (for the original virus that circulated in 2020, i.e., prior to its replacement by more transmissible variants) and CV$\approx 1$. The gap in $\mathcal{R}_0$ is evidently due to the lack of interventions in this simple model, lack of account for under-reporting of cases, and many other features that have been well studied in the COVID-19 pandemic. Potentially more interesting is the comparison of the CV estimates. While \citep{Gomes2022} invested in estimating the entire individual variation subject to selective depletion, here we consider a single index that not only is a collection of several factors -- income, employment, health and disability, education, housing, crime, living environment -- but also represents an average over data zones. The present estimate can, therefore, be interpreted as explaining roughly $5\% $ of the total individual variation estimated in \citep{Gomes2022}. 

Future work might consist in increasing the granularity of these models, but this will require more data. It would be informative, for example, to see the socioeconomic index broken down into its constituent factors, the size of the data zones ($784$ individuals on average) reduced (ideally down to individual level), as well as to seek additional data on factors that are currently not represented in SIMD. These might include age, contact patterns, and biological markers reflecting susceptibility to infection. To infer more risk factors by the inverse problem approach described here, the epidemic metapopulation model would need to be extended to more patches to ensure solvability. We expect the gap between the two CV estimates to reduce as models and data are further refined. Meanwhile, a similar approach has been initiated for endemic diseases, such as tuberculosis \citep{Gomes2019} and malaria \citep{Corder2020}.

\section*{Acknowledgements}
This work is partially funded by FCT – Fundação para a Ciência e a Tecnologia, I.P., under the scope of the projects UIDB/00297/2020 (https://doi.org/10.54499/UIDB/00297/2020) and UIDP/00297/2020 (https://doi.org/10.54499/UIDP/00297/2020) (Center for Mathematics and Applications).
 
\bibliography{references}

\end{document}